# AI-aided multiscale modeling of physiologically-significant blood clots

Yicong Zhu[a], Changnian Han[a], Peng Zhang[a], Guojing Cong[b], James R.Kozloski[c], Chih-Chieh Yang[c], Leili Zhang[c] and Yuefan Deng[a,] *

[a] Department of Applied Mathematics and Statistics, Stony Brook University, Stony Brook, NY, USA
[b] Oak Ridge National Laboratory, Oak Ridge, TN, USA
[c] IBM Thomas J. Watson Research Center, Yorktown Heights, NY, USA

**Abstract:**

We have developed an AI-aided multiple time stepping (AI-MTS) algorithm and multiscale modeling framework (AI-MSM) and implemented them on the Summit-like supercomputer, AIMOS. AI-MSM is the first of its kind to integrate multi-physics, including intra-platelet, inter-platelet, and fluid-platelet interactions, into one system. It has simulated a record-setting multiscale blood clotting model of 102 million particles, of which 70 flowing and 180 aggregating platelets, under dissipative particle dynamics to coarse-grained molecular dynamics. By adaptively adjusting timestep sizes to match the characteristic time scales of the underlying dynamics, AI-MTS optimally balances speeds and accuracies of the simulations.

1. Introduction

Cardiovascular diseases (CVD) are the leading threat to human life, estimated to account for 17.9 million deaths each year by the World Health Organization (WTO), or 32% of deaths globally. The most prominent cause of fatalities among CVD are due to heart attack or stroke, and are largely related to thrombosis formation [1]. Meanwhile, flow-induced platelet activation can be evoked by a high shear stress environment [2]. Therefore, the increasing probability of platelet activation and thrombotic risk occurring under stenotic flow in implanted blood-recirculating devices is also a life-threating phenomenon, widely discussed among researchers recently [3]. Many clinical studies of COVID-19 report markedly abnormal coagulation in severe patients, while platelet-rich thrombi lead to dysfunction across multiple organs [4, 5]. Enormous efforts have been made to simulate thrombus formation using different numerical methods and help researchers understand the biochemical mechanism and mechanical stimuli underlying the different stages of thrombosis formation.

Thrombosis formation is a complex physiological process including several important modules, such as fluid mechanics, platelet activation, and platelet aggregation [6]. Differing spatial and temporal scales of these fundamental modules require that researchers develop multiscale computational methods. In this study, we developed a multiscale particle-based biomechanical model, which consists of 250 platelets in shear blood flows made up of 102 million particles. This model represents a record number of particles on HPC used to simulate thrombosis formation at molecular resolution.

In this study, we applied a multi-scale modelling (MSM) framework that integrates the DPD-CGMD methods to study the dynamics and mechanical properties of thrombosis formation [7, 8]. The dissipative particle dynamics (DPD) is used to model the hydrodynamics of blood flow and the coarse-grained molecular dynamics (CGMD) is used to model the platelet particles. When commonly implemented with conventional single time stepping (STS), MSM integrates the dynamics at the finest time resolution to ensure sufficient system-wide accuracy, but at the expense of massively redundant computations that over-resolve larger-scale phenomena. In these situations, the STS requires significant resources to compute high temporal resolutions that might be unnecessary. Therefore, researchers [9] introduced several multiple time stepping (MTS) algorithms to split the interactions based on the characteristic spatial scales of different types of interactions and assign different timestep sizes to each scale.

To address the vast spatial and temporal scales in this multi-component system, we developed a novel artificial intelligence (AI) algorithm that adapts multiple timestep sizes to the underlying biophysical dynamics. This algorithm improves the overall simulation speed without causing significant loss of modeling accuracy. This AI-aided multiple time stepping (AI-MTS) algorithm, general in principle, is implemented specifically to study blood clotting on the AiMOS. The simulated system is decomposed based on the different scales of biological components and the associated timestep sizes is predicted on the fly by an AI algorithm. Among the many components, the AI-MTS algorithm consists of two recurrent neural network (RNN) autoencoders (AEs) that predict the timestep sizes and number of steps where the size is adjusted by analyzing the time series of platelet dynamics and extracting the latent features. The AI-MTS algorithm is evaluated by examining the generated mechanics and thermodynamics and their computational loads. In summary, our main contributions are:

1. **The application**: We modeled a large blood clot of 250 platelets in shear blood flows consisting of 102 million particles, the first of such efforts to integrate multi-physics at multi-scales, serving as a numerical system for studying platelet dynamics and characterizing mechanotransduction of hemodynamic stresses, at the cellular level.
2. **The algorithm**: We designed an AI inference system that adaptively adjusts timestep sizes on the fly to accommodate the varying dynamics of platelet-involved events occurring at various temporal scales. The AI-MTS significantly accelerates simulation while generating sufficiently accurate dynamical details.
3. **The implementation**: We implemented the algorithms of inter-platelet and platelet-fluid interactions and the AI-MTS integrator on AiMOS. AI-MTS can reduce the simulation time from 91 years to 3 months for a 1 millisecond simulated time, greatly enhanced the feasibility of modeling physiologically-significant biomedical systems with the currently available HPC resources.

The rest of the manuscript is organized as follows: Section 2 outlines the physiology and mathematical models of blood clotting to set the stage for and motivate our entire study. Section 3 addresses the computational aspects of multiscale modeling with focus on the AI-MTS algorithm. Section 4 describes the implementation, and analyzes performance, of the AI-MTS algorithm. Section 5 summarizes the impacts and extensions of AI-MTS for its generalization and improvement for blood-clotting simulation.

**2. The physiology and model of blood clotting**

Understanding life-threatening thrombus formation is the focus of many computational studies and is paramount to understanding the risks related to COVID-19 and many CVD. In our previous study, we mainly focused on one important process, platelet aggregation, wherein free-flowing platelets' resting membrane binds fibrinogen to initiate thrombi [10, 11]. This phenomenon may end with so-called "white clots", or platelet-rich thrombi, which can then travel to other sites and prevent blood flow [12]. In our current study, we created a typical stenosis flow environment by significantly extending the simulation size to create a more realistic representation of thrombosis formation in human blood, which typically consists of hundreds of platelets.

Since numerous complex components, such as blood flow, blood cells, receptors and ligands are involved in thrombosis growth, an extreme need exists for a well-organized and developed framework at multiple scales. The all-atom molecular dynamics (MD) simulation is the standard and finest method to capture highly relevant details during simulation, and is widely applied for nanoscale particle-based mechanics studies. However, MD simulation is restricted by its computational costs, and therefore presents a serious and unavoidable challenge when integrated into simulations at nanometers in length and nanoseconds in time scales. Another popular particle-based simulation method, the coarse-grained molecular dynamics (CGMD) method [13], fuses groups of particles as beads. As a result of the reduction of computed particles, CGMD is less computationally intensive and can be used in simulations at larger spatial and temporal scales (reaching micrometers and milliseconds). As a mesoscopic simulation technique, the dissipative particle dynamics (DPD) [14] uses a set of particles to represent the molecular fragments and fluid regions. All particles interact by three forces: conservative, dissipative and random forces. This is also a popular choice for fluid simulations at the scales of millimeters and milliseconds. In addition, the traditional continuum method, computational fluid dynamics (CFD), which uses equations to analyze and govern the fluid flows, is also used frequently in many numerical studies. Although these different methods are designed for their corresponding characteristic scales, a complex millisecond-scale process such as ours, which couples multiple length scales into a single system is still a major development and computational challenge. Thus observing through simulation the meaningful dynamics of blood clotting at the molecular scale currently requires prohibitively long simulations, and this challenge has therefore motivated our current study.

In recent decades, many approaches have been developed to model different phenomena that occur in thrombus formation process. Pivkin *et al* [15] utilized force coupling method (FCM), to model the formation of platelet thrombi. By introducing activation delay time, they demonstrated how thrombus growth rate is influenced by blood velocity. Yazdani *et al* [16] further applied the FCM method and correlated the adhesion force to local shear rates to establish a shear-dependent model of platelets adhesion over a wide range of flow shear rate. Mody and King [17, 18] developed a 3D platelet-aggregation model, Platelet Adhesive Dynamics (PAD), to study interplatelet collisions and high shear-induced transient aggregation via GPIbα-vWF binding. They used Monte Carlo method to determine the bonding profile and CDL-BIEM code to predict particle-wall hydrodynamic interactions. Then, based on their PAD method, Wang *et al* [19] simulated the tethering of platelets on the injured vessel surface mediated by GPIba-vWF and analyzed the interaction between flowing platelets and adherent platelets. Tosenberger *et al* [20, 21] provided a hybrid DPD-PDE model to study the role of platelets in consecutive stages of blood coagulation. They used DPD to model plasma and platelets and PDEs to model the concentration of biochemical substances.

In each of these studies, continuum methods have failed to capture the specific binding dynamics and mechanical details at the molecular level. Furthermore, pure atomistic MD is unrealistic to simulate the milliseconds of flow-induced mechanotransduction in vessel required to understand this phenomenon. In this study, we aim to encapsulate a few hundred platelets to simulate physiologically realistic blood clots and study them at fine molecular scales in space and time. Therefore, we developed a combined DPD-CGMD multiscale model and captured the molecular features to describe mechanical details that occur on each platelet as well as other cellular biophysical events of interest, e.g., platelet aggregation ranging over many spatial and temporal scales.

In summary, the joint capabilities of the MSM framework and the newly developed AI-MTS and their implementation on state-of-the-art supercomputers enable us to simulate a record number of platelets for record time scales, making the simulation physiologically significant. This study, while still at an early stage, has already revealed key mechanisms of blood clotting and thereby is poised for contribution to therapeutic advances including drug discovery [22].

## 3. The computational solutions of the model

### 3.1. The blood clotting problem

The AI-MSM framework in this work integrates multi-physics including intra-platelet, inter-platelet, and fluid-platelet interactions into one system for the first time, which is adapted to simulate platelet-mediated blood clotting with 102 million particles, spreading in 70 flowing and 180 aggregating platelets under shear blood flow (Figure 1). A microchannel is simulated by a rectangular cross-section with dimensions of $x \times y \times z = 302.3 \times 21.3 \times 19.9$ in μm with periodic boundary conditions in $x$- and $z$-dimensions. The blood vessel walls are modeled at top and bottom boundaries, where the no-slip boundary condition is applied. The pressure-driven Poiseuille flow mimicking shear blood flow moves in the positive $x$-direction.

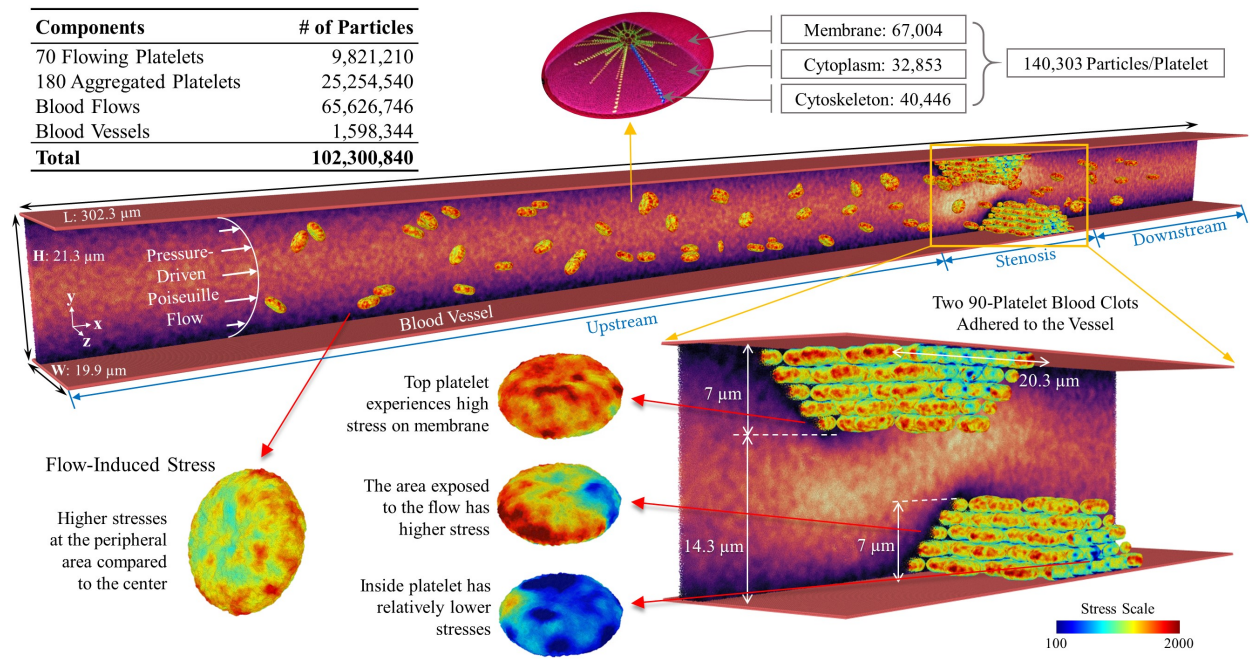

**Figure 1.** The simulated blood clotting system of 102,300,840 particles, involving 70 flowing platelets and 2 adhered clots, each with 90 aggregated platelets. A slice of fluid is colored by the flow speed. Two blood clots of 7 μm in size, forming a 14.6 μm stenosis. Three platelets from the clot and one flowing platelet, as examples, are colored in membrane flow-induced stresses.

Inside the flow region, 70 randomly distributed flowing platelets were initiated and two artificial blood clots, each of 90 platelets, adhered to the walls, form a stenosis of 14.3 μm in height. We decomposed the system into three sections by flow conditions: upstream section, stenosis, and downstream section. The entire system contains 102,300,840 particles, incorporating 65,626,746 fluid particles, 1,598,344 vessel wall particles and 250×140,303 = 35,075,750 particles for the 250 platelets. The NVE ensemble is applied to such a neutral system in the simulation. Using a sheer size of 102 million particles to characterize and explore blood clotting, including platelet rotation, adhesion, and aggregation at the millisecond scale, is unprecedented and was therefore only made possible by the tactical models, orchestrated efforts of machine learning, MSM algorithms, HPC architecture, and methodical algorithm implementations that we report here.

### 3.2. Multiscale solution method

We describe the platelet model and its complexities in correlating multiple spatiotemporal scales. The MSM of platelet under shear blood flow (Figure 1) covers two major scales: a mesoscopic scale for the viscous flow modeled by DPD and the microscopic scale for the platelet constituents, including membrane, cytoplasm, and cytoskeleton modeled by CGMD [7, 8, 11]. Each DPD particle represents an assemblage of related atoms, and their collective motion is governed by

$$d\mathbf{v}_i = \frac{1}{m_i}\sum_{j\neq i}\left(\mathbf{F}_{ij}^C dt + \mathbf{F}_{ij}^D dt + \mathbf{F}_{ij}^R \sqrt{dt} + \mathbf{F}^E dt\right) \tag{1}$$

$$\mathbf{F}_{ij}^C = \begin{cases} \alpha\left(1-\frac{r_{ij}}{r_c}\right)\mathbf{e}_{ij}, & r_{ij} < r_c \\ 0, & r_{ij} \geq r_c \end{cases},$$

$$\mathbf{F}_{ij}^D = -\gamma \omega^D(r_{ij})(\mathbf{e}_{ij}\cdot\mathbf{v}_{ij})\mathbf{e}_{ij},$$

$$\mathbf{F}_{ij}^R = \sigma \omega^R(r_{ij})\zeta_{ij}\sqrt{dt}\,\mathbf{e}_{ij},$$

where $\mathbf{F}_{ij}^C, \mathbf{F}_{ij}^D, \mathbf{F}_{ij}^R, \mathbf{F}_{ij}^E$ are conservative, dissipative, random, and external forces. $r_{ij}$ is the inter-particle distance, $\mathbf{v}_{ij}$ is the relative velocity, $\mathbf{e}_{ij}$ is the unit vector in the direction $r_{ij}$, $\xi_{ij}$ is a random variable following normal distribution, and parameters $\alpha, \gamma$, and $\sigma$ control force strength [14, 23]. The intra-platelet interaction is governed by

$$V(r_{ij}) = \sum_{\substack{\text{nonbonded}\\j\neq i}}\left(U_{LJ}(r_{ij}) + U_M(r_{ij})\right) + \sum_{\text{bonds}} k_b(r_{ij}-r_0)^2 + \sum_{\text{angles}} k_\theta(\theta_{ij}-\theta_0)^2, \tag{2}$$

$$U_{LJ}(r) = 4\varepsilon\left[\left(\frac{\sigma}{r}\right)^{12} - \left(\frac{\sigma}{r}\right)^{6}\right],$$

$$U_M(r) = \varepsilon\left[e^{\alpha(1-r/r_0)} - 2e^{\alpha(1-r/r_0)/2}\right],$$

where $U_{LJ}$ and $U_M$ indicate the Lennard-Jones (LJ) and Morse potentials and the following three terms are bonded interactions with force constants $k_b, k_\theta$, and $k_\phi$ [7]. The platelet-fluid interface is described by a hybrid force field

$$d\mathbf{v}_i = \frac{1}{m_i}\sum_{j\neq i}\left(\nabla U_{LJ}(r_{ij})dt + \mathbf{F}_{ij}^D dt + \mathbf{F}_{ij}^R \sqrt{dt}\right). \tag{3}$$

where the LJ potential to maintain the incompressibility of the platelet and the stochastic and random terms preserve thermodynamics and mechanical properties [7]. The inter-platelet interaction is modeled by combining the Morse potential to mimic the long-medium range effects of aggregation with a harmonic term to bind receptors [8]

$$d\boldsymbol{v}_i = \frac{1}{m_i} \sum_{j \neq i} \left( \nabla U_M + f^A (1 - r_{ij}/r_0) dt \right). \tag{4}$$

Figure 2 demonstrates the MSM's schematic details spanning the orders of magnitudes of 6 in space and 9 in time. A pressure-driven Poiseuille flow is applied to simulate shear blood flow and induce platelet rotation and aggregation.

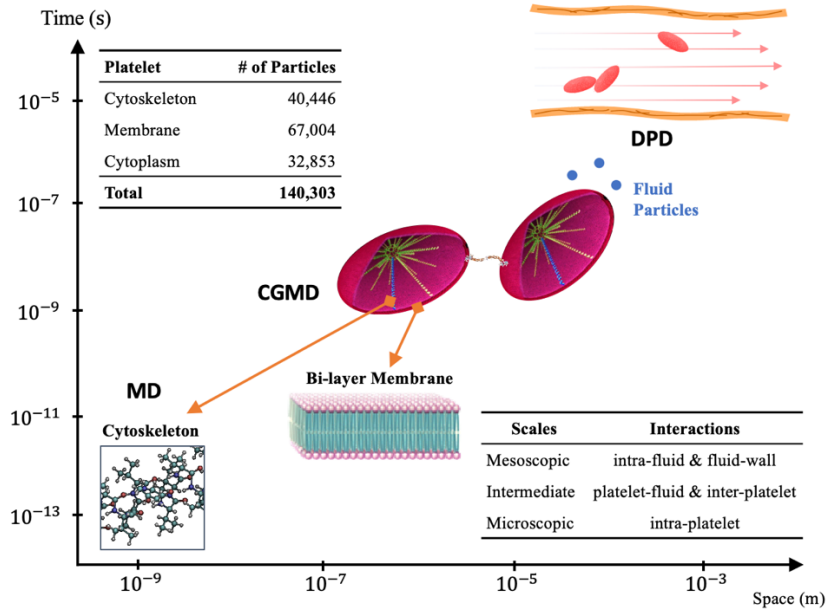

**Figure 2. The multiscale models of platelet and associated force fields.**

### 3.3. The AI-MTS algorithm

We introduce our AI-aided simulation algorithms in this section. Conventional MTS algorithms such as RESPA [9], developed for single-scale MD simulations, split force calculations by interaction ranges. For more complex problems including blood clotting, multiple force types spanning 6 spatial and 9 temporal scales require component splitting that segregates the system based on components' characteristic spatial scales [24, 25]. For intricate and highly dynamical systems of complex biophysical events, such as platelet-involved rotation, aggregation and adhesion while interacting with fluid or other platelets, the lack of adaptivity to varying temporal scales in simulation algorithms introduces great implementation complexity to examining dynamics and adjusting timestep sizes during simulation. Since simulations at all scales proceed concurrently, the timestep sizes need to be adjusted seamlessly to small and large for high- and low-frequency dynamics, respectively, for optimal efficiency. The timestep sizes regulation on the fly for a sophisticated multi-physics system can benefit from AI techniques with automated system states analysis. We proposed an AI-enhanced Adaptive Time Stepping (AI-ATS) algorithm that automatically analyzes platelet dynamics and adapts time resolutions to the underlying states [26], where platelet dynamics is analyzed by neural networks.

We then developed an AI-MTS algorithm integrating a dynamics-analysis AI system with a component-splitting MTS algorithm to automatically accommodate the multiple spatiotemporal scales in our MSM for speeding up integrations. Our MTS algorithm splits the system into two subsystems based on the spatial scales of the underlying components: the fluid at the macroscopic scale and the platelet at the microscopic scale. A larger static timestep size in the range of microseconds to nanoseconds is applied to the fluid hemodynamics, and a smaller adaptive timestep size at picoseconds to nanoseconds is applied to the platelet to capture molecular-level dynamics. To adapt timestep sizes, AI is invoked to analyze the platelet dynamics and to predict the time resolution on the fly.

**Table 1. The pseudocode of the MSM simulator.**

| | |
|---|---|
| 1 | loop over N steps with $\Delta t_f$: |
| 2 |   update $\mathbf{v}(\Delta t_f/2)$ for fluid particles |
| 3 |   neighbor lists construction and communication |
| 4 |   loop over $k_m$ steps with $\Delta t_m = \Delta t_f/k_m$: |
| 5 |     update $\mathbf{v}(\delta t_1/2)$ for flowing platelets at this level |
| 6 |     loop over $k_2$ steps $\delta t_2 = \delta t_1/k_2$: |
| 7 |       ⋮ |
| 8 |       loop over $k_c$ steps with $\delta t_c = \delta t_{c-1}/k_c$: |
| 9 |         update $\mathbf{v}(\delta t_c/2)$ for aggregating platelets |
| 10 |         update $\mathbf{x}(\delta t_c)$ for all particles |
| 11 |         communication |
| 12 |         compute forces for aggregating platelets |
| 13 |         update $\mathbf{v}(\delta t_c)$ |
| 14 |       ⋮ |
| 15 |       compute forces for flowing platelets |
| 16 |       update $\mathbf{v}(\delta t_1)$ |
| 17 |     compute forces for fluid-platelet interface |
| 18 |     update $\mathbf{v}(\delta t_m)$ |
| 19 |   compute forces for fluid |
| 20 |   update $\mathbf{v}(\delta t_f)$ |
| 21 |   compute and dump platelet physical properties |
| 22 |   $t \leftarrow t + \Delta t_f$, counter $s \leftarrow s + 1$ |
| 23 |   if $s = l$ then |
| 24 |     trigger AI inference |
| 25 |     synchronize inference results |

**Table 2. The AI inference pipeline.**

| | |
|---|---|
| 1 | import raw data $X$ in time-series format |
| 2 | denoised data $\tilde{X} \leftarrow \text{filter}(X)$ |
| 3 | latent features $X_L \leftarrow \text{AE}(\tilde{X})$ |
| 4 | $\left[\Delta t_p^{(\text{new})}, l\right] \leftarrow \text{FCN}(X_L)$ |
| 5 | categorize $\Delta t_p^{(\text{new})}$ to integration levels |
| 6 | export platelets' id and corresponding levels and $l$ |

A three-level MTS time integrator is developed to handle the two subsystems and the interface using different timestep sizes. The characteristic $\Delta t_f$ is employed to integrate the fluid subsystem aggressively, and a variable $\Delta t_p$ attuning the dynamics is applied to the platelet subsystem for capturing molecular-level details while an intermediate $\Delta t_m$, where $\Delta t_f > \Delta t_m \geq \Delta t_p$, handles the fluid-platelet interface bridging the energy exchange in between. A factor $k$ indicating the ratio of timestep sizes between levels helps balance the time resolutions. The platelet subsystem is further divided into sublevels for the low-resolution flowing platelet motion and the high-resolution inter-platelet aggregation formation. The pseudocode (Table 1) of our MSM simulator summarizes our description of the platelet dynamics and the characterization of the platelet motion by six physics measurements, calculated during the simulation, in time series including angular velocity components, kinetic energies, and the surrounding flow speed at the platelet center of mass. AI captures the dynamics with three modules: (a) two-stage denoising filters including moving average (MA) and wavelet transform (WT) for the high-frequency oscillation reduction, generating denoised time series; (b) two-layer LSTM/GPU-based autoencoders (AEs) (16 and 4 neuron units in layer 1 and 2, respectively) taking the most recent 50 data points of 200 μs long as the input and producing 8 representative latent features and finally (c) 4-layer feedforward neural networks (FCNs) (32, 16, 8, 1 neurons per layer) that determine the optimal timestep sizes under current dynamics, as illustrated in Table 2.

The training process follows [26]. The training dataset was composed by simulations of flowing platelets under shear blood. The raw time-evolving physics measurements were collected as time series and denoised by MA. By further splitting, each training sample is formatted as a 2-D matrix of dimensions 50 × 6 (steps × features) and a total of 1,600 samples. The labeling of timestep sizes is based on a trail-and-error approach to determine the most optimal one by examining possible values. Since the $\Delta t_f$ must be multiples of $\Delta t_p$, the label is categorized in terms of the factor $k$. Same activation and loss functions as in [26] were applied. AEs and FCNs were trained separately, where AEs act as latent feature extractors and FCNs as predictors.

## 4. Implementation and Performance Analysis

In this section, we introduce the HPC architecture, the implementation of our AI-aided algorithms and performance analyses in details. For the HPC architecture, we present the details of AiMOS, the Summit-like heterogeneous supercomputer. For the implementation of AI-MTS algorithms, we present the workflow for intelligently matching modeling time resolutions with the dynamics. In analyzing performances, we compare the AI-MTS over STS in terms of their accuracies and speeds.

### 4.1. Implementation

AiMOS supercomputer, ranking 48th and 17th in the June 2021 releases of the TOP500 and Green500 lists, consists of 252 compute nodes each containing 2 IBM POWER 9 processors at 3.15 GHz. Each processor has 20 cores with 4 hardware threads per core, running with 256 GB DDR4 memory of 135 GB/s peak bandwidth. The aggregated performance is 0.97 TFlop/s for 2 CPUs. All compute nodes are interconnected with the Mellanox EDR InfiniBand and connected to the unified GPFS file system.

We implemented the hybrid force fields of platelet-involved interactions and the AI-MTS integrator by encapsulating LAMMPS (version 29Oct2020) [27]: (a) three LAMMPS pair forces were developed to handle inter- and intra-platelet interactions; (b) a three-level time integrator schema was implemented to incorporate integration levels, from the top-level fluid to the mid-level fluid-platelet interface to the bottom-level platelets, and the bottom-level was further decomposed into sublevels for accommodating adaptive timestep sizes; (c) the

`PairHybrid::compute()` was modified for invoking component-based force calculations according to integration levels; (d) at each level, the `initial_integrate()` and `final_integrate()` were modified to take `mts_level` as an extra parameter, indicating the integration level where the current component belongs, and update positions and velocities accordingly; (e) the neural-network architecture in the AI inference system was built and attached to the end of the time integration.

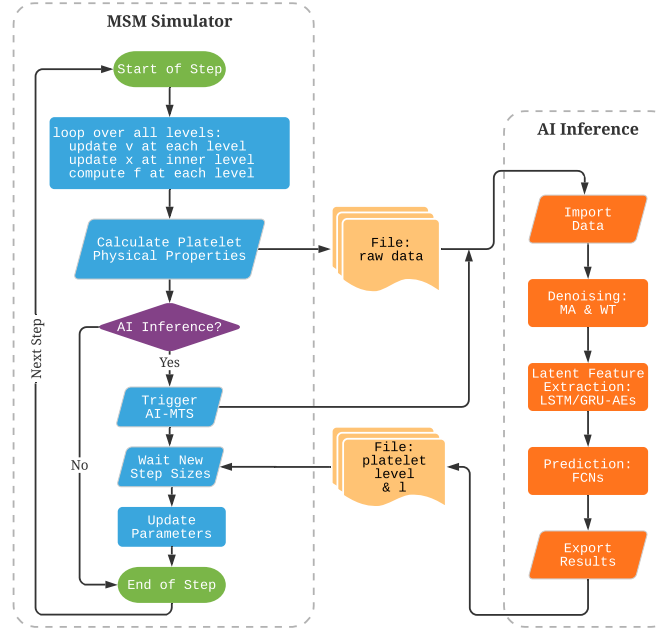

Figure 3. The AI-MTS workflow.

Table 3. The software environments of AiMOS.

| Module | AiMOS |
| --- | --- |
| MPI | IBM Spectrum MPI 10.3.0.01 |
| Host Compiler | GCC 4.8.5 |
| TensorFlow | TensorFlow 2.1.1 |

Figure 3 depicts the workflow of the AI-MTS algorithm in one integration step. At the start of time integration, the MSM simulator carried out the three-level time integration and calculated the target physics measurements, which are exported to files in the file system as the raw and noisy data. Once the AI inference is invoked, it imports, and processes, raw data through the inference pipeline for denoising and predicting the timestep sizes. The inference outputs are transferred back to the MSM simulator through the file system. After making necessary parameter synchronizations, the simulation resumes with new timestep sizes.

Both STS and AI-MTS algorithms were implemented for simulation speed comparison. We used 36 MPI tasks per node for each experiment. In STS, the static step size is applied to the whole system following [24], acting as the reference system. In AI-MTS, the platelet $\Delta t_p$ varies among 52/104/208/416 ps, where the highest resolution is required to capture inter-platelet interactions in the aggregation and low resolutions are matched to flowing platelets with different physical states. The choice of $\Delta t_p$ is determined by the AI inference on the platelet dynamics on the fly. The fluid system at the top scale is advanced with a large static $\Delta t_f$ = 1.664 ns since the flow is driven by a constant pressure. The same force parameters in [8] are used in simulations. All simulations were conducted by our modified LAMMPS package, integrated with the AI program using TensorFlow. The software environment is listed in Table 3.

### 4.2. Performance Analysis

The accuracy performance of the AI-MTS algorithm is analyzed against the STS algorithm. The system thermodynamics was expressed by the temperature and pressure. The flow condition was examined by the speed profile. The platelet mechanics was investigated by the flow-induced stress on the membrane. The number of floating-point operations (FLOPs) per time step is 135 GFLOPs.

We first studied the accuracy performance of the AI-MTS algorithm on our MSM. The system temperature and pressure were monitored on a 0.5 μs basis. Table 4 summarizes the average and deviation by AI-MTS and STS in a 50 μs period. AI-MTS preserves system thermodynamics with 0.7% and 1.2% deviation in temperature and pressure from STS. The developed flow speed in the x-direction and fluid particle density distribution at different positions in the y-axis for both algorithms are compared in Figure 4, presenting a consistent parabolic flow pattern.

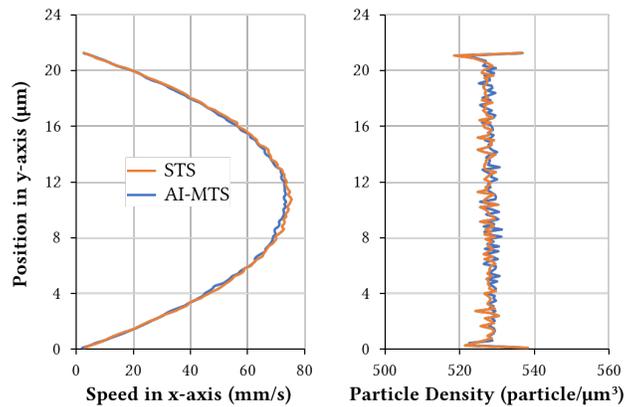

Figure 4. The flow speeds and densities vs. the position in y-axis from two algorithms.

Table 4. The accuracy analysis of STS and AI-MTS.

|  | STS | AI-MTS | Deviation |
|---|---|---|---|
| Temperature | 1.145±0.001 | 1.152±0.001 | 0.7% |
| Pressure | 146.42±1.75 | 148.18±2.06 | 1.2% |

The blood clots in the simulation consists of several layers: the outermost shell-layer interfacing the flow, the layers beneath it and the inner permeable core. As a result, the shell-layer platelets are exposed to the highest shear stresses, decreasing towards the inner-layers. Figure 5 selected four representative platelets as examples, one flowing platelet and three aggregated platelets from the clot, to demonstrate the platelet mechanics by the flow-induced stress on the platelet membrane. For the aggregated platelets from the clot, we observed higher stress on those platelets located at the clot surface exposed to the flow, compared to those hidden inside the clot that have fewer interactions with the fluid. In addition, for the flowing platelet, higher stress is observed on the peripheral area compared to the center area.

The stress profile on the membrane presents identical patterns in both AI-MTS and STS results. The relative deviation of stress is computed by the particle-wise stress differences between two algorithms, yielding at most 3% deviation. In summary, the system, flow, and platelet properties are well maintained in AI-MTS simulations.

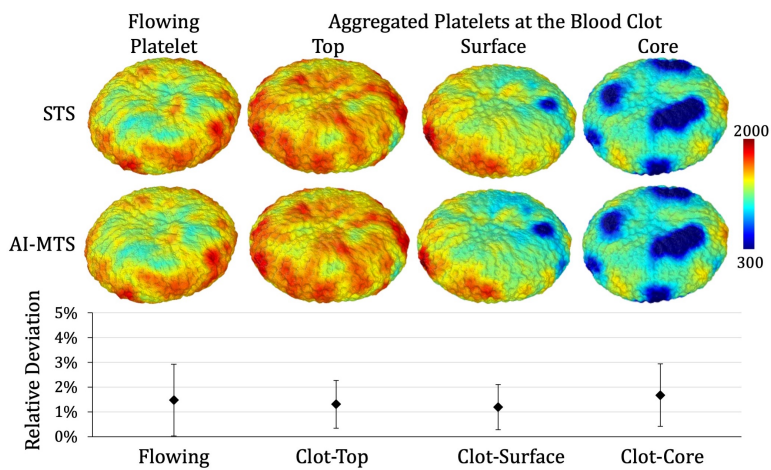

**Figure 5. A visual and quantitative comparison of the distributed stresses on the membrane obtained by two algorithms.**

**Table 5. The simulation speeds of STS and AI-MTS algorithms (unit: μs/day).**

| Number Of Nodes | STS | AI-MTS | Speedup ratio (AI-MTS/STS) |
|---|---|---|---|
| 32 | 0.02 | 3.60 | 194 |
| 64 | 0.03 | 8.78 | 282 |

For speed analysis, the AI-MTS algorithm reduces the computing intensity by adapting time resolutions to biology components and their dynamics. A common practice to benchmark in speed performance is to estimate the ratio of simulated time to simulating time, following the Gordon Bell winning reports of Anton-2 [28]. We present the performance results of STS and AI-MTS algorithms on AiMOS in Table 5. The speed results show that: (1) the speed of STS is less than 0.03 μs/day; (2) Benchmarking against STS, AI-MTS achieves consistent simulation speed improvements on CPUs, with a speedup ratio up to 194 on 32 nodes, and 282 on 64 nodes. In summary, compared to STS, AI-MTS achieves great simulation speed improvements on AiMOS without significantly impacting the accuracy of the simulated dynamics and mechanics properties.

## 5. Discussion and Conclusions

We modeled a record application of multi-physics MSM of platelet-mediated blood clotting under shear blood flow involving 102 million particles to characterize complex biological processes spanning 6 spatial scales and 9 temporal scales. We developed an AI-MTS algorithm to optimize computations automatically and intelligently for

MSM. Specifically, the system is split into 3 subsystems by biological components and timestep sizes are adapted to varying dynamics by a deep learning-based AI inference system on-the-fly. We implemented a multi-level time integrator and benchmarked our proposed algorithm against STS on AiMOS for accuracy and speed analysis. Our AI-MTS algorithm proved its superiority in accelerating the complex simulation, with a speedup ratio of up to 282, and its capability of preserving the mechanics and thermodynamics features with over 97% accuracy. The performance demonstrates that the AI-MTS algorithm reduces the computing time for 1 millisecond simulation from 91 years to 3 months with 64 computing nodes, enabling studies of nearly realistic clotting of millisecond durations at nanometer scales of detail. An MSM can efficiently characterize and model complex biological processes spanning a wide range of scales, by the orchestrated efforts of algorithms including adaptive space and time resolutions, machine learning and their implementation on the latest HPC architectures, for holistic optimization.

Our continuing efforts with intelligent blood clotting models will include optimization for AI-MTS workflow for given HPC architectures and the development of AI inference for maximizing resource utilization. We propose as next steps for this research: (1) Implementing GPU-enabled pair force calculations. We managed to compute the three CPU-version pair forces to handle inter-and intra-platelet interactions through LAMMPS. With GPU utilization, the current blood clotting simulation on AiMOS, configured with IMB POWER9 CPUs and NVIDIA Volta V100 GPUs, will be further significantly accelerated. (2) A new domain decomposition method. Load imbalance is a big challenge for complex bioscience models, since multiple components and multiple interactions at different scales are included. Given the implicit nature of the blood clotting system, while the density of blood clots and their surrounding fluid varies as much as 100-fold, problems can arise in terms of loads of computation, communication, and memory utilization, etc. A domain decomposition-based load balancer is required to reduce the disparity in particle numbers across computing units and thus shorten the occurring overhead and boost the overall simulation speed. (3) Fusing CPU-multithreading. The current simulator could be further enhanced by fusing CPU-multithreading and optimize the intra-node communication for better node-wise performance. (4) Improving the synchronization overhead. The demanding frequent data movement between the host and accelerators further strains the precious bandwidth, which needs to be addressed in the future.

**Acknowledgements**

This project is supported by the SUNY-IBM Consortium Award, IPDyna: Intelligent Platelet Dynamics, FP00004096 (PI: Y. Deng). The simulations in this study were conducted on the AiMOS supercomputer at Rensselaer Polytechnic Institute and the SeaWulf Cluster at Stony Brook University (PIs: Y. Deng and P. Zhang).